\RequirePackage{ifpdf}
\ifpdf 
\documentclass[pdftex]{sigma}
\else
\documentclass{sigma}
\fi

\begin{document}


\renewcommand{\PaperNumber}{101}

\FirstPageHeading

\ShortArticleName{On a Whitham-Type Equation}

\ArticleName{On a Whitham-Type Equation}

\Author{Sergei SAKOVICH}

\AuthorNameForHeading{S. Sakovich}

\Address{Institute of Physics, National Academy of Sciences, 220072 Minsk, Belarus}
\Email{\href{mailto:saks@tut.by}{saks@tut.by}}

\ArticleDates{Received September 27, 2009, in f\/inal form November 05, 2009;  Published online November 08, 2009}

\Abstract{The Hunter--Saxton equation and the Gurevich--Zybin system are considered as two mutually non-equivalent representations of one and the same Whitham-type equation, and all their common solutions are obtained exactly.}

\Keywords{nonlinear PDEs; transformations; general solutions}

\Classification{35Q58; 35C05}

\section{Introduction}

The following Whitham-type equation
\begin{gather}
u_t = 2 u u_x - \partial_x^{-1} u_x^2 \label{e1}
\end{gather}
was proposed recently by Prykarpatsky and Prytula in \cite{PP} as a model equation describing the short-wave perturbations in an abstract elastic one-dimensional medium with relaxation and spatial memory ef\/fects.

This equation \eqref{e1}, containing the ill-def\/ined term $\partial_x^{-1} u_x^2$, was represented in \cite{PP} by the second-order nonlinear partial dif\/ferential equation
\begin{gather}
u_{xt} = 2 u u_{xx} + u_x^2 , \label{e2}
\end{gather}
and it was shown there that \eqref{e2} is integrable in the sense of possessing a bi-Hamiltonian structure, an inf\/inite hierarchy of conservation laws, and a Lax-type representation. Also, two f\/inite-dimensional reductions of \eqref{e2} were obtained in \cite{PP}, and they turned out to be some integrable by quadratures dynamical system, which could be useful for deriving wide classes of exact solutions of \eqref{e2}.

A dif\/ferent representation of \eqref{e1}, the hydrodynamic-type system
\begin{gather}
u_t = 2 u u_x - v , \qquad v_t = 2 u v_x , \label{e3}
\end{gather}
was proposed recently by Bogoliubov, Prykarpatsky, Gucwa, and Golenia in \cite{BPGG}. In the f\/irst equation of \eqref{e3}, the ill-def\/ined term $\partial_x^{-1} u_x^2$ of \eqref{e1} was replaced by the new variable $v$, while the time evolution of $v$ was determined by an additional f\/irst-order equation, the second equation of~\eqref{e3}. It was announced in \cite{BPGG} that \eqref{e3} is integrable in the sense of possessing a Lax-type representation. The system \eqref{e3} was called in \cite{BPGG} an integrable regularization of the Whitham-type equation \eqref{e1}.

In the present paper, we consider the nonlinear PDEs \eqref{e2} and \eqref{e3} in Sections \ref{s2} and \ref{s3}, respectively. We give the references which show that these equations \eqref{e2} and \eqref{e3}, especially the f\/irst one of them, were known and quite well studied in the literature prior to \cite{PP,BPGG}. We put a particular emphasis on the fact that the equations \eqref{e2} and \eqref{e3} can be completely solved by quadratures, and we ref\/ine the derivations of their general solutions, consistently following the way used for the Rabelo equations in \cite{SS}. We have the following reasons to re-derive the general solutions of \eqref{e2} and \eqref{e3}. Firstly, the dif\/ferent derivations of the general solutions of \eqref{e2}, given in the literature, all overlooked the evident class of $x$-independent solutions. Secondly, the general solution of \eqref{e3} can be expressed not only in a parametric form, known in the literature, but also in an implicit form. And thirdly, we need to get the results in uniform notations, for Section~\ref{s4}, where we make a comparison of the general solutions of \eqref{e2} and \eqref{e3}, taking into account that these two representations of the Whitham-type equation \eqref{e1} are not equivalent to each other, and obtain all common solutions of \eqref{e2} and \eqref{e3} exactly.

\section[The Hunter-Saxton equation]{The Hunter--Saxton equation}
\label{s2}

The nonlinear equation \eqref{e2} is far not novel. Up to a scale transformation of its variables, this is the celebrated Hunter--Saxton equation \cite{HS}, sometimes referred to as the Hunter--Zheng equation \cite{HZ1}. The Hunter--Saxton equation has been studied in almost all respects, including its complete solvability by quadratures \cite{HS,DP,P1,M}, relationship with the Camassa--Holm equation and the Liouville equation \cite{DP,P1}, bi-Hamiltonian formulation \cite{HZ1,OR}, integrable f\/inite-dimensional reductions \cite{HZ1,BSS}, global solution properties \cite{HZ2,BC}, and geometric interpretations \cite{R,L}, to mention only a few of numerous publications on this equation.

\looseness=1
In our opinion, the most important feature of the Hunter--Saxton equation is the possibility to obtain its general solution in a closed form. This equation is linearizable \cite{M}, or C-integrable in the Calogero's terminology, but it also belongs to a subset of C-integrable equations whose general solutions can be expressed in a closed form. Such completely sol\-vable equations of the Liouville equation's type dif\/fer from other C-integrable equations of the Burgers equation's type, and from the so-called S-integrable (completely integrable, or Lax integrable) equations of the sine-Gordon equation's type, in many respects. For example, the Liouville equation possesses a~continuum of variational symmetries (hence, a continuum of nontrivial conservation laws) and several Lax-type representations which all turn out to be equivalent to conservation laws \cite{S}. Leaving a study of such properties of the Hunter--Saxton equation for a separate publication, here we only concentrate on its general solution. The general solution of \eqref{e2} can be obtained in a~parametric form, in at least three dif\/ferent ways \cite{HS,P1,M}. The derivation we give below is similar to the original one of~\cite{HS}, but dif\/fers from it by a more precise treatment of the arbitrariness of the transformation involved, in the spirit of~\cite{SS}.

Making the transformation
\begin{gather}
x = x(y,t) , \quad x_y \neq 0 , \qquad u(x,t) = a(y,t) , \label{e4}
\end{gather}
where the function $x(y,t)$ is initially not f\/ixed, and using the identities
\begin{gather}
u_x = \frac{a_y}{x_y} , \qquad u_t = a_t - \frac{a_y x_t}{x_y} , \qquad u_{xx} = \frac{a_{yy}}{x_y^2} - \frac{a_y x_{yy}}{x_y^3} , \notag \\
u_{xt} = \frac{a_{yt}}{x_y} - \frac{a_{yy} x_t + a_y x_{yt}}{x_y^2} + \frac{a_y x_{yy} x_t}{x_y^3} , \label{e5}
\end{gather}
we bring the nonlinear equation \eqref{e2} into the form
\begin{gather}
a_{yt} + \frac{a_y^2}{x_y} = \partial_y \left( \frac{(x_t + 2 a) a_y}{x_y} \right) . \label{e6}
\end{gather}
Now we see from \eqref{e6} that it is expedient to f\/ix the function $x(y,t)$ of the transformation \eqref{e4} by the condition $x_t + 2 a = 0$, which brings the equation into a constant-characteristic form and considerably simplif\/ies it. Doing this, we f\/ind that the transformation \eqref{e4} with
\begin{gather}
a = - \tfrac{1}{2} x_t \label{e7}
\end{gather}
relates the second-order equation \eqref{e2} with the third-order equation
\begin{gather}
x_{ytt} - \frac{x_{yt}^2}{2 x_y} = 0 \label{e8}
\end{gather}
which follows from \eqref{e6} and \eqref{e7}.

Through the transformation \eqref{e4} with \eqref{e7}, the general solution of the third-order equation \eqref{e8} represents the general solution of the second-order equation \eqref{e2} parametrically, with $y$ being the parameter. Note, however, that, according to the Cauchy--Kovalevskaya theorem \cite{O}, the ge\-ne\-ral solution of \eqref{e8} must contain three arbitrary functions of one variable, whereas the general solution of \eqref{e2} must contain only two arbitrary functions of one variable. This redundant arbitrariness in $x(y,t)$, caused by the invariance of \eqref{e8} with respect to an arbitrary transformation $y \mapsto Y(y)$ which has no ef\/fect on $u(x,t)$ of \eqref{e2}, can be eliminated by the following normalization of the parameter $y$. We rewrite \eqref{e8} in the form
\begin{gather*}
\partial_t \big( x_y^{-1/2} x_{yt} \big) = 0 , 
\end{gather*}
integrate over $t$, and get
\begin{gather}
x_y^{-1/2} x_{yt} = f(y) , \label{e10}
\end{gather}
where $f(y)$ is an arbitrary function. For any nonzero function $f(y)$, we can set, without loss of generality, $f = 2$ in \eqref{e10} by an appropriate transformation $y \mapsto Y(y)$ which does not change the corresponding solutions of \eqref{e2}, where the value $2$ is chosen for convenience only. The case of $f = 0$ must be considered separately. Consequently, all solutions of the second-order equation~\eqref{e2} are represented parametrically by all solutions of the second-order equation \eqref{e10} with $f = 0$ and $f = 2$ through the transformation \eqref{e4} with \eqref{e7}.

The case of $f = 0$ in \eqref{e10} is $x_{yt} = 0$, which immediately leads us through \eqref{e7} and \eqref{e5} to $a_y = 0$ and $u_x = 0$, that is, to the evident class of solutions
\begin{gather}
u = \tau(t) \label{e11}
\end{gather}
of \eqref{e2}, with any function $\tau(t)$. In the case of $f = 2$, we integrate \eqref{e10} over $t$ and get
\begin{gather}
x_y = \bigl( t + \phi(y) \bigr)^2 , \label{e12}
\end{gather}
with any function $\phi(y)$. Then, integrating \eqref{e12} over $y$ and using \eqref{e7} and \eqref{e4}, we obtain the following class of solutions of \eqref{e2}, determined parametrically:
\begin{gather}
x = y t^2 + 2 t \int \phi(y) \, dy + \int \phi(y)^2 \, dy + \psi(t) , \notag \\
u(x,t) = - y t - \int \phi(y) \, dy - \frac{1}{2} \psi'(t) , \label{e13}
\end{gather}
where $y$ is the parameter, $\phi(y)$ and $\psi(t)$ are arbitrary functions, and the prime denotes the derivative. The expressions \eqref{e11} and \eqref{e13} together constitute the general solution of the second-order nonlinear partial dif\/ferential equation \eqref{e2}.

Some words are due on the obtained general solution of \eqref{e2}. It follows from \eqref{e13} that
\begin{gather}
u_x = \frac{-1}{t + \phi(y)} . \label{e14}
\end{gather}
According to this relation \eqref{e14}, the condition $u_x \neq 0$ is satisf\/ied for any function $\phi(y)$, which proves that the class of solutions \eqref{e13} does not cover solutions of the class \eqref{e11}. For some unknown reasons, only the parametric expressions \eqref{e13} were called the general solution of \eqref{e2} in \cite{HS,P1,M}, whereas the solutions \eqref{e11} were omitted there. Also, the relation \eqref{e14} makes clear that all solutions of \eqref{e2}, except for those of the class \eqref{e11}, inevitably possess singularities of the type $u_x = \pm \infty$, when considered on the interval $- \infty < t < \infty$. The transformation \eqref{e4}, used for obtaining the general solution of \eqref{e2}, is applicable everywhere outside those singularities $u_x = \pm \infty$, because the condition $x_y \neq 0$ is satisf\/ied due to \eqref{e12}. This inevitable presence of singularities in the solutions \eqref{e13} was noticed in \cite{HS}. In the next section, we show that nontrivial solutions of the representation \eqref{e3} of the Whitham-type equation \eqref{e1} not necessarily contain blow-ups of derivatives.

\section[The Gurevich-Zybin system]{The Gurevich--Zybin system}
\label{s3}

Proceeding to the hydrodynamic-type system \eqref{e3}, we note that this is the one-dimensional reduction of the Gurevich--Zybin system \cite{GZ1,GZ2} which can be completely solved by quadratures~\cite{GZ2,P2}. For an earlier appearance of \eqref{e3} in plasma physics, one can consult Section 3 of~\cite{D}. In~\cite{BD}, a bi-Hamiltonian structure and a zero-curvature representation were found and studied for the system \eqref{e3}. Below we show how to obtain the general solution of~\eqref{e3} in an implicit form, following the way used in \cite{SS}.

Applying the transformation
\begin{gather}
x = x(y,t) , \quad x_y \neq 0 , \qquad
u(x,t) = a(y,t) , \qquad v(x,t) = b(y,t) \label{e15}
\end{gather}
to the system \eqref{e3}, with $x(y,t)$ being not f\/ixed initially, we obtain
\begin{gather}
a_t - \frac{(x_t + 2 a) a_y}{x_y} + b = 0 , \qquad b_t - \frac{(x_t + 2 a) b_y}{x_y} = 0 . \label{e16}
\end{gather}
Then we f\/ix the function $x(y,t)$ in \eqref{e15} and \eqref{e16} by the condition $x_t + 2 a = 0$, and thus get
\begin{gather*}
a = - \tfrac{1}{2} x_t , \qquad b = \tfrac{1}{2} x_{tt} , \qquad x_{ttt} = 0 , 
\end{gather*}
that is,
\begin{gather}
x = \alpha(y) t^2 + \beta(y) t + \gamma(y) , \qquad
a = - \alpha(y) t - \tfrac{1}{2} \beta(y) , \qquad b = \alpha(y) , \label{e18}
\end{gather}
where $\alpha(y) , \beta(y) , \gamma(y)$ are three arbitrary functions, of which at least one is non-constant due to $x_y \neq 0$. The expressions \eqref{e15} and \eqref{e18} represent the general solution of the system \eqref{e3} parametrically, with $y$ being the parameter. An appropriate transformation $y \mapsto Y(y)$, which has no ef\/fect on solutions $u(x,t)$, $v(x,t)$ of \eqref{e3}, may be used to f\/ix any one of the three arbitrary functions in \eqref{e18}.

This parametric general solution of \eqref{e3} can be expressed in an implicit form, as follows. When the function $\alpha(y)$ is non-constant, we replace $a(y,t)$ and $b(y,t)$ in \eqref{e18} by $u(x,t)$ and $v(x,t)$, respectively, then eliminate $y$ from the resulting expressions, and thus obtain
\begin{gather}
x + v t^2 + 2 u t + \mu(v) = 0 , \qquad u + v t + \nu(v) = 0 , \label{e19}
\end{gather}
where $\mu(v)$ and $\nu(v)$ are arbitrary functions (expressible in terms of the arbitrary functions~$\alpha$, $\beta$, $\gamma$). When $\alpha(y)$ is constant but $\beta(y)$ is not, we do the same and get
\begin{gather}
x + \xi t^2 + 2 u t + \rho(u + \xi t) = 0 , \qquad v = \xi , \label{e20}
\end{gather}
where $\rho$ is an arbitrary function of its argument, and $\xi$ is an arbitrary constant. When $\alpha(y)$ and~$\beta(y)$ are constant but $\gamma(y)$ is not, we get
\begin{gather}
u = \eta t + \zeta , \qquad v = - \eta , \label{e21}
\end{gather}
where $\eta$ and $\zeta$ are arbitrary constants. These expressions \eqref{e19}--\eqref{e21} together constitute the general solution of the nonlinear system~\eqref{e3}.

Unlike all nontrivial solutions of the Hunter--Saxton equation, some solutions of the Gurevich--Zybin system \eqref{e3}, of the class \eqref{e19}, do not contain blow-ups of derivatives. Indeed, it follows from \eqref{e19} that the expression for any derivative of $u$ or $v$ contains only some degree of the expression $t^2 + 2 t \nu'(v) - \mu'(v)$ in its denominator, for example,
\begin{gather*}
u_x = \frac{- t - \nu'(v)}{t^2 + 2 t \nu'(v) - \mu'(v)} , \qquad v_x = \frac{1}{t^2 + 2 t \nu'(v) - \mu'(v)} , 
\end{gather*}
where the prime denotes the derivative. Clearly, it is possible to choose the functions $\mu$ and $\nu$ so that $t^2 + 2 t \nu'(v) - \mu'(v) \neq 0$ holds on the whole interval $- \infty < t < \infty$.

\section{Discussion}
\label{s4}

The general solutions of the Hunter--Saxton equation \eqref{e2} and the Gurevich--Zybin system \eqref{e3} are quite dif\/ferent in their structure. The general solution of~\eqref{e2} is given in the parametric form \eqref{e13}, except for the explicit solutions~\eqref{e11}. The general solution of~\eqref{e3} is given in the implicit form~\eqref{e19} and~\eqref{e20}, except for the explicit solutions \eqref{e21}. From this point of view, the Hunter--Saxton equation \eqref{e2} and the Gurevich--Zybin system~\eqref{e3} are very similar to the exp-Rabelo equation $u_{xt} = \exp u - ( \exp u )_{xx}$ and the quadratic Rabelo equation $u_{xt} = 1 + \frac{1}{2} ( u^2 )_{xx}$, respectively \cite{SS}.

The nonlinear PDEs \eqref{e2} and \eqref{e3} were considered in \cite{PP,BPGG} as two well-def\/ined representations of the Whitham-type equation \eqref{e1} which itself contains the ill-def\/ined term $\partial_x^{-1} u_x^2$. Evidently, these two representations are not equivalent to each other. The Gurevich--Zybin system \eqref{e3} can be re-written as the second-order equation
\begin{gather}
u_{tt} - 4 u u_{xt} + 4 u^2 u_{xx} - 2 u_x u_t + 4 u u_x^2 = 0 \label{e23}
\end{gather}
for $u(x,t)$ with the def\/inition $v = 2 u u_x - u_t$ for $v(x,t)$, and this second-order equation \eqref{e23} dif\/fers from the Hunter--Saxton equation \eqref{e2}. For this reason, one may wonder whether the PDEs \eqref{e2} and \eqref{e3} have any common nontrivial solutions at all.

It can be found easily that the compatibility condition for the equations \eqref{e2} and \eqref{e23} is
\begin{gather}
u_{tt} = 4 u^2 u_{xx} + 2 u_x u_t . \label{e24}
\end{gather}
Alternatively, in the variables $u$ and $v$, the compatibility condition for the PDEs \eqref{e2} and \eqref{e3} is
\begin{gather}
v_x = u_x^2 . \label{e25}
\end{gather}
One can f\/ind all common solutions of \eqref{e2} and \eqref{e3} by applying the condition \eqref{e24} to the ge\-neral solution \eqref{e11} and \eqref{e13} of the Hunter--Saxton equation, or, alternatively, by applying the condition~\eqref{e25} to the general solution \eqref{e19}--\eqref{e21} of the Gurevich--Zybin system. Using the condition~\eqref{e24}, we get
\begin{gather}
\tau'' = 0 \label{e26}
\end{gather}
from \eqref{e11}, and
\begin{gather}
\psi''' = 0 \label{e27}
\end{gather}
from \eqref{e13}. Using the condition \eqref{e25}, we get
\begin{gather}
\mu' + \nu'^2 = 0 \label{e28}
\end{gather}
from \eqref{e19}, while \eqref{e20} does not satisfy \eqref{e25}, and \eqref{e21} satisf\/ies \eqref{e25} identically. It is quite obvious that \eqref{e11} with \eqref{e26} is equivalent to \eqref{e21}, and that \eqref{e13} with \eqref{e27} is equivalent to \eqref{e19} with \eqref{e28}.

Thus, summarizing the result in a nonrigourous way, we can say that the degree of arbitrariness of common nontrivial solutions of the Hunter--Saxton equation \eqref{e2} and the Gurevich--Zybin system \eqref{e3} is one arbitrary function of one variable.

\subsection*{Acknowledgement}

The author is deeply grateful to Professor E.V.~Ferapontov and Professor M.V.~Pavlov for pointing out the origin of the system~\eqref{e3}, to the referees for their useful suggestions, and to the Max-Planck-Institut f\"{u}r Mathematik for hospitality and support.

\pdfbookmark[1]{References}{ref}
\LastPageEnding


\begin{thebibliography}{99}

\footnotesize\itemsep=1pt

\bibitem{PP} Prykarpatsky A.K., Prytula M.M.,
The gradient-holonomic integrability analysis of a Whitham-type nonlinear dynamical model for a relaxing medium with spatial memory,
{\it Nonlinearity} {\bf 19} (2006), 2115--2122.

\bibitem{BPGG} Bogoliubov N.N. Jr., Prykarpatsky A.K., Gucwa I., Golenia J., Analytical properties of an Ostrovsky--Whitham type dynamical system for a relaxing medium with spatial memory and its integrable regularization, \href{http://arxiv.org/abs/0902.4395}{arXiv:0902.4395}.

\bibitem{SS}
Sakovich A., Sakovich S.,
On transformations of the Rabelo equations, {\it SIGMA} {\bf 3} (2007), 086, 8 pages,
\href{http://arxiv.org/abs/0705.2889}{arXiv:0705.2889}.

\bibitem{HS}
Hunter J.K., Saxton R.,
Dynamics of director f\/ields,
{\it SIAM J. Appl.\ Math.} {\bf 51} (1991), 1498--1521.

\bibitem{HZ1}
Hunter J.K., Zheng Y.,
On a completely integrable nonlinear hyperbolic variational equation,
{\it Phys. D} {\bf 79} (1994), 361--386.

\bibitem{DP}
Dai H.-H., Pavlov M.,
Transformations for the Camassa--Holm equation, its high-frequency limit and the Sinh-Gordon equation,
{\it J. Phys.\ Soc.\ Japan} {\bf 67} (1998), 3655--3657.

\bibitem{P1}
Pavlov M.V.,
The Calogero equation and Liouville-type equations,
{\it Theoret. and Math.\ Phys.} {\bf 128} (2001), 927--932,
\href{http://arxiv.org/abs/nlin.SI/0101034}{nlin.SI/0101034}.

\bibitem{M}
Morozov O.I.,
Contact equivalence of the generalized Hunter--Saxton equation and the Euler--Poisson equation, \href{http://arxiv.org/abs/math-ph/0406016}{math-ph/0406016}.

\bibitem{OR}
Olver P.J., Rosenau P.,
Tri-Hamiltonian duality between solitons and solitary-wave solutions having compact support,
{\it Phys.\ Rev.\ E} {\bf 53} (1996), 1900--1906.

\bibitem{BSS}
Beals R., Sattinger D.H., Szmigielski J.,
Inverse scattering solutions of the Hunter--Saxton equation,
{\it Appl.\ Anal.} {\bf 78} (2001), 255--269.

\bibitem{HZ2}
Hunter J.K., Zheng Y.X.,
On a nonlinear hyperbolic variational equation. I.~Global existence of weak solutions,
{\it Arch. Rational Mech. Anal.} {\bf 129} (1995), 305--353.

\bibitem{BC}
Bressan A., Constantin A.,
Global solutions of the Hunter--Saxton equation,
{\it SIAM J. Math.\ Anal.} {\bf 37} (2005), 996--1026,
\href{http://arxiv.org/abs/math.AP/0502059}{math.AP/0502059}.

\bibitem{R}
Reyes E.G.,
The soliton content of the Camassa--Holm and Hunter--Saxton equations,
in Proceedinds of Fifth International Conference ``Symmetry in
Nonlinear Mathematical Physics'' (July 9--15, 2001, Kyiv),
Editors A.G.~Nikitin, V.M.~Boyko and R.O.~Popovych,
{\it Proceedings of Institute
of Mathematics, Kyiv} {\bf 43} (2002), Part~1, 201--208.

\bibitem{L}
Lenells J.,
The Hunter--Saxton equation: a geometric approach,
{\it SIAM J. Math.\ Anal.} {\bf 40} (2008), 266--277.

\bibitem{S}
Sakovich S.Yu.,
On conservation laws and zero-curvature representations of the Liouville equation,
{\it J.~Phys.~A: Math.\ Gen.} {\bf 27} (1994), L125--L129.

\bibitem{O}
Olver P.J.,
Applications of Lie groups to dif\/ferential equations, 2nd ed., {\it Graduate Texts in Mathematics}, Vol.~107, Springer-Verlag, New York, 1993.

\bibitem{GZ1}
Gurevich A.V., Zybin K.P.,
Nondissipative gravitational turbulence,
{\it Soviet Phys.\ JETP} {\bf 67} (1988), 1--12.

\bibitem{GZ2}
Gurevich A.V., Zybin K.P.,
Large-scale structure of the Universe. Analytic theory,
{\it Soviet Phys.\ Usp.} {\bf 38} (1995), 687--722.

\bibitem{P2}
Pavlov M.V.,
The Gurevich--Zybin system,
{\it J. Phys.~A: Math.\ Gen.} {\bf 38} (2005), 3823--3840,
\href{http://arxiv.org/abs/nlin.SI/0412072}{nlin.SI/0412072}.

\bibitem{D}
Davidson R.C.,
Methods in nonlinear plasma theory, Academic Press, New York, 1972.

\bibitem{BD}
Brunelli J.C., Das A.,
On an integrable hierarchy derived from the isentropic gas dynamics,
 {\it J. Math.\ Phys.} {\bf 45} (2004), 2633--2645,
\href{http://arxiv.org/abs/nlin.SI/0401009}{nlin.SI/0401009}.

\end{thebibliography}
\end{document}